\documentclass[namedreferenced]{kluwer}
\usepackage{graphicx}

\begin{document}


\newcommand{\apj}{{\em Ap. J.}} 	    
\newcommand{\apjl}{{\em Ap. J. (Letters)}}  

\newcommand{\chisq}{\mbox{$\chi^2$ }}	    
\newcommand{\chinu}{\mbox{$\chi_{\nu}^2$}}  
\newcommand{\degrees}{\mbox{$^{\circ}$}}    
\newcommand{\degree}{\mbox{$^{\circ}$}}    
\newcommand{\degdot}{\mbox{$. \! ^{\circ}$}} 
\newcommand{\etal}{\mbox{\it et al.\ }}
\newcommand{\flux}{\mbox{erg cm$^{-2}$ s$^{-1}$}}
\newcommand{\Ka}{\mbox{K$\alpha$}}
\newcommand{\NH}{\mbox{${\rm N_H}$ }}	    
\newcommand{\NHunits}{\mbox{$\times 10^{20} {\rm cm}^{-2}$}}

\newcommand{\Meszaros}{M\'{e}sz\'{a}ros}    

\newcommand{\astrod}{{\it ASCA}}
\newcommand{\asca}{{\it ASCA}}
\newcommand{\ASCA}{{\it ASCA}}
\newcommand{\asuka}{{\it ASCA}}
\newcommand{\acis}{{\it ACIS}}
\newcommand{\BeppoSAX}{{\it BeppoSAX}}
\newcommand{\axaf}{{\it Chandra}}
\newcommand{\chandra}{{\it Chandra}}
\newcommand{\cubic}{{\it CUBIC}}
\newcommand{\CUBIC}{{\it CUBIC}}
\newcommand{\EXOSAT}{{\it EXOSAT}}
\newcommand{\rosat}{{\it ROSAT}}
\newcommand{\RXTE}{{\it RXTE}}
\newcommand{\sacb}{{\it SAC-B}}
\newcommand{\swift}{{\it Swift}}
\newcommand{\Swift}{{\it Swift}}
\newcommand{\Tenma}{{\it Tenma}}
\newcommand{\xmm}{{\it XMM-Newton}}
\newcommand{\XMM}{{\it XMM-Newton}}
\newcommand{\XRTDAS}{{\it XRTDAS}}

\begin{article}

\begin{opening}

\title{THE SWIFT X-RAY TELESCOPE}

\author{David N. \surname{Burrows}}
\author{J. E. \surname{Hill}}
\author{J. A. \surname{Nousek}}
\author{J. A. \surname{Kennea}}
\institute{Pennsylvania State University, 525 Davey Lab, University
  Park, PA 16802, USA
 \email{burrows@astro.psu.edu}}

\author{A. \surname{Wells}}
\author{J. P. \surname{Osborne}} 
\author{A. F. \surname{Abbey}} 
\author{A. \surname{Beardmore}}
\author{K. \surname{Mukerjee}}
\author{A.D.T. \surname{Short}}
\institute{Space Research Centre, University of Leicester, Leicester LE1 7RH, UK}

\author{G. \surname{Chincarini}} 
\author{S. \surname{Campana}}
\author{O. \surname{Citterio}} 
\author{A. \surname{Moretti}}
\author{C. \surname{Pagani}}
\author{G. \surname{Tagliaferri}} 
\institute{INAF-Osservatorio Astronomico di Brera, Via Bianchi 46,
23807 Merate, Italy}

\author{P. \surname{Giommi}}
\author{M. \surname{Capalbi}}
\author{F. \surname{Tamburelli}} 
\institute{ASI Science Data Center, via Galileo Galilei, 00044 Frascati, Italy}

\author{L. \surname{Angelini}}
\institute{NASA/Goddard Space Flight Center, Greenbelt, MD 20071, USA}

\author{G. \surname{Cusumano}}
\institute{INAF-Istituto di Astrofisica Spaziale e Fisica Cosmica Sezione di Palermo,  
                 Via Ugo La Malfa 153, 90146 Palermo, Italy}

\author{H. W. \surname{Br\"{a}uninger}} 
\author{W. \surname{Burkert}}
\author{G. D. \surname{Hartner}} 
\institute{Max-Planck-Institut f\"{u}r Extraterrestrische Physik, Garching
  bei M\"{u}nchen, Germany}

\runningtitle{The {\Swift} X-ray Telescope}
\runningauthor{D. N. Burrows et al.}

\abbreviations{
\abbrev{BAT}{Burst Alert Telescope};
\abbrev{CCD}{Charge-Coupled Device};
\abbrev{GRB}{Gamma-Ray Burst};
\abbrev{HPD}{Half-Power Diameter};
\abbrev{PSF}{Point Spread Function};
\abbrev{TAM}{Telescope Alignment Monitor};
\abbrev{TEC}{Thermo-Electric Cooler};
\abbrev{TDRSS}{Tracking and Data Relay Satellite System};
\abbrev{UVOT}{Ultra-Violet/Optical Telescope};
\abbrev{XRT}{X-Ray Telescope}
}

\dedication{Dedicated to David J. Watson, in memory of his valuable
  contributions to this instrument.}


\begin{ao}
David Burrows,
2582 Gateway Dr.,
State College, PA 16801  USA
\end{ao}

\begin{abstract}
The \Swift\ Gamma-Ray Explorer is designed to make prompt
multiwavelength observations of Gamma-Ray Bursts (GRBs) and GRB
afterglows.  The X-ray Telescope (XRT) enables
\Swift\ to determine GRB positions with a few arcseconds accuracy
within 100 seconds of the burst onset.  

The XRT utilizes a mirror set built for JET-X and an
\XMM/EPIC MOS  CCD detector to provide a sensitive
broad-band (0.2-10 keV) X-ray imager with effective area of $> 120$ cm$^2$ at
1.5 keV, field of view of 23.6 x 23.6 arcminutes, and angular
resolution of 18 arcseconds (HPD).  The detection sensitivity is
2x10$^{-14}$ erg cm$^{-2}$ s$^{-1}$ in 10$^4$ seconds.  The instrument is designed to
provide automated source detection and position reporting within 5
seconds of target acquisition.  It can also measure the redshifts of GRBs
with Fe line emission or other spectral features.  The XRT
operates in an auto-exposure mode, adjusting the CCD readout mode
automatically to optimize the science return for each frame as the
source intensity fades.  The XRT will measure spectra and lightcurves of the GRB
afterglow beginning about a minute after the burst and will follow
each burst for days or weeks.
\end{abstract}


\end{opening}

\section{INTRODUCTION}

The \Swift\ Gamma Ray Burst Explorer \cite{gehrels2004} was chosen in October 1999 as NASA's
next MIDEX mission, and was launched on 20 November 2004.  It
carries three instruments: a Burst Alert Telescope (BAT; \opencite{6}), which
identifies gamma-ray bursts (GRBs) and determines their location on
the sky to within a few arcminutes; an Ultraviolet/Optical Telescope
(UVOT; \opencite{7}) with limiting sensitivity of 24$^{\rm{th}}$
magnitude in 1000 s and with 0.3 arcsecond
position accuracy; and an X-ray Telescope (XRT).  The three instruments
combine to make a powerful multiwavelength observatory with the
capability of rapid position determinations of GRBs to arcsecond
accuracy within 1-2 minutes of their discovery, and the ability to
measure both lightcurves and redshifts of the bursts and afterglows.

\begin{figure}[h]
\centerline{\includegraphics[bb=85 104 509 690,clip=true,width=2.5in,angle=270]{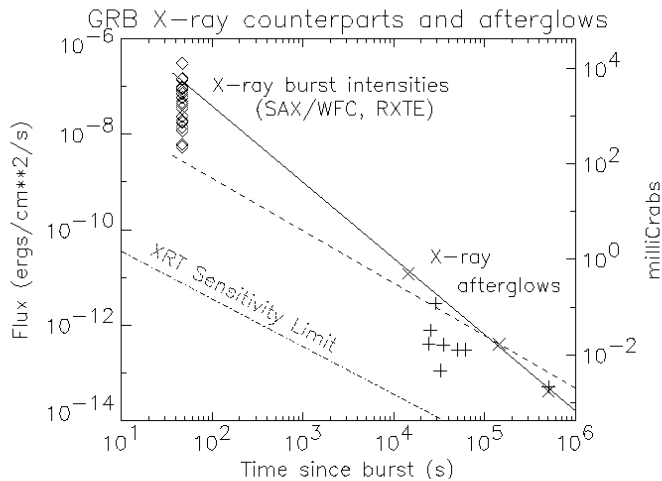}}
\caption{X-ray burst and afterglow lightcurves.  The cluster of data
  points at about 1 minute represents peak X-ray burst fluxes for a
  number of bursts observed in 1997-1999 by {\BeppoSAX} and \RXTE.  The data points
  between $10^4$ and $10^6$ s after the burst are \BeppoSAX, \RXTE, and
  {\ASCA} afterglow flux measurements for a subset of these bursts.  
  Note the data gap of about $10^4$
  seconds, during which the afterglow flux drops by about 4 orders of
  magnitude.  The \Swift\ XRT will fill in this data gap.}
\label{fig:decay}
\end{figure}

The \Swift\ XRT is a sensitive, flexible, autonomous X-ray imaging
spectrometer designed to measure fluxes, spectra, and lightcurves of
GRBs and afterglows over a wide dynamic range of more than 7 orders of
magnitude in flux.  The \BeppoSAX\ satellite showed that accurate GRB
positions can be effectively determined by a good X-ray telescope,
since over 84\% of GRBs have X-ray afterglows \cite{12}, compared with about
50\% for optical afterglows. However, by the time that {\BeppoSAX} was
able to observe a typical X-ray afterglow, its intensity had already
dropped by 4-5 orders of magnitude (Figure~\ref{fig:decay}). 

The \Swift\ XRT will begin
observations before the GRB ends in some cases, and will fill in the
large time gap during which the Lorentz factor of the relativistic
blast wave changes from $\sim 100$ to $<10$.  It will provide accurate
positions within 5 seconds of target acquisition for typical bursts,
allowing ground-based optical telescopes to begin immediate
spectroscopic observations of the afterglow.

We previously reported on the instrument design at an early stage in
its development \cite{15}.  Here we describe the final design and performance
of the XRT.

\section{OVERALL DESCRIPTION}

\begin{figure}[h]
\centering
\includegraphics[height=\maxfloatwidth,angle=270]{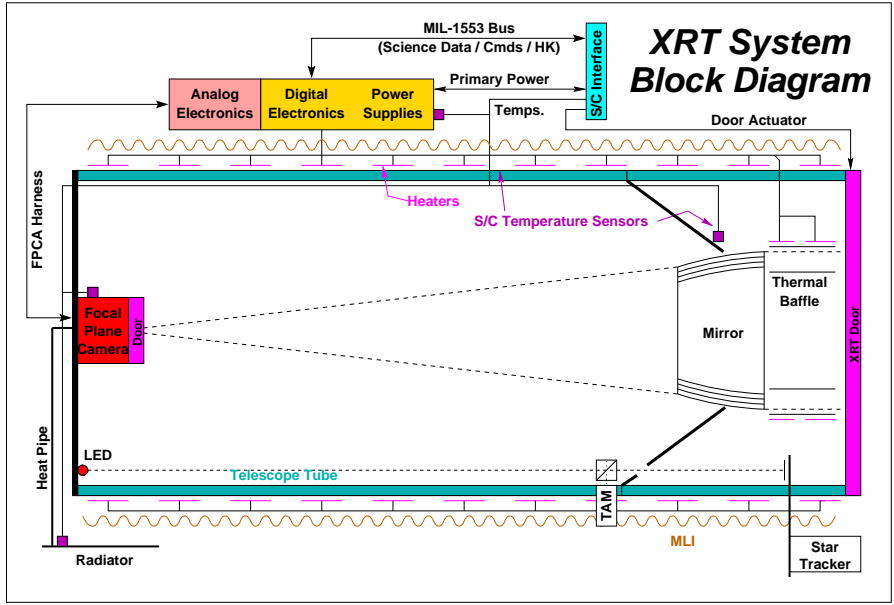}
\caption{Block diagram of the XRT.  The instrument design is described
in detail in \S\ref{sec:design}.  The TAM is the Telescope Alignment Monitor, an
internal alignment monitor described in \S\ref{sec:TAM}.}
\label{fig:layout1}
\end{figure}

\begin{figure}
\centering
\includegraphics[bb=121 117 462 707,clip=true,height=\maxfloatwidth,angle=270]{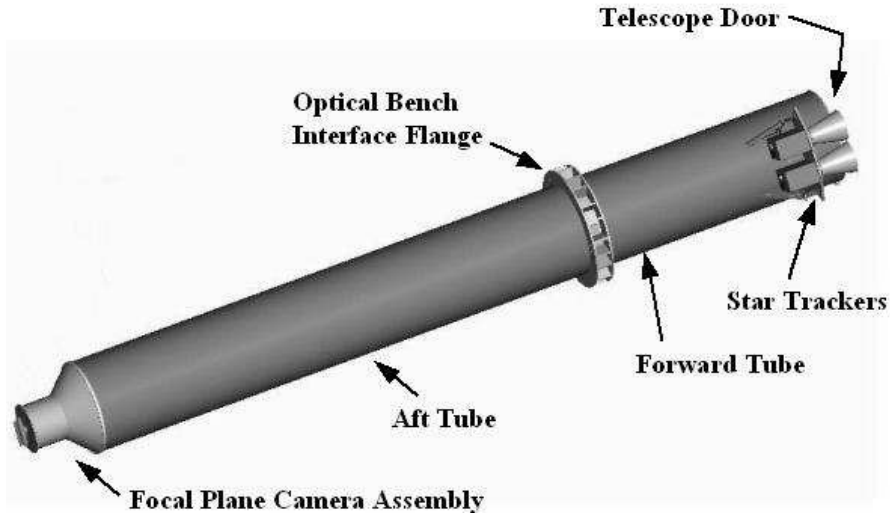}
\caption{XRT design.  The telescope focal length is 3.500m, 
and the overall instrument length is 4.67m, with a diameter of 
0.51m.}
\label{fig:layout2}
\end{figure}

\begin{figure}
\centering
\includegraphics[width=3.0in,angle=90]{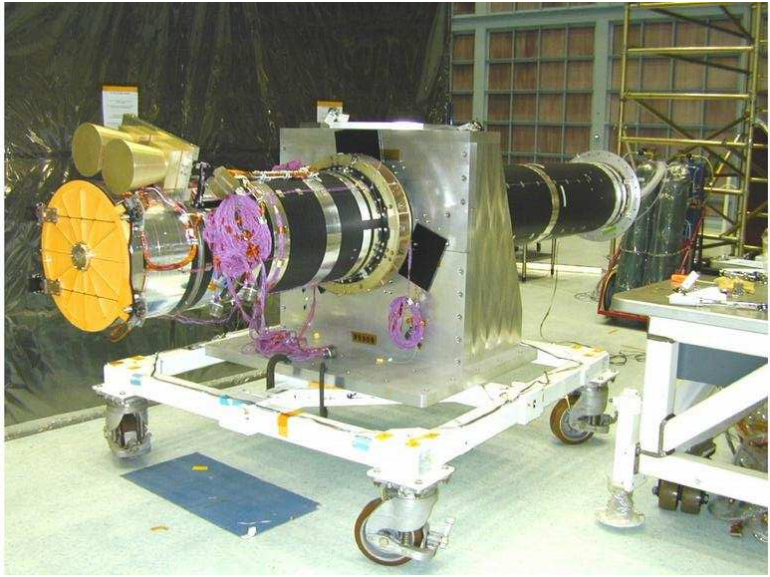}
\caption{XRT installed in test cart before thermal vacuum tests in July
  2002, with star tracker simulators installed.
The telescope door is at the left end.  The silver bands around the
tube are aluminum tape over the operational and survival heaters.}
\label{fig:photo1}
\end{figure}

The XRT uses a grazing incidence Wolter I telescope to focus X-rays
onto a thermoelectrically cooled CCD.  The layout of the instrument is shown in
Figures~\ref{fig:layout1} and \ref{fig:layout2}.  
Figure~\ref{fig:photo1} shows the completed instrument
(before thermal blanket installation).  A door protects the mirrors during launch.
A thermal baffle provides a warm environment for the front end of the
mirrors to prevent thermal gradients in the mirror module that could distort
the Point Spread Function (PSF).  The mirrors are the JET-X
flight spares (\opencite{Citterio96}; \opencite{2}; \opencite{8}), 
and were calibrated at the Panter X-ray Calibration
Facility of the Max-Planck-Institute f\"{u}r Extraterrestrische Physik in
1996 \cite{Citterio96} and again in July 2000.  A composite telescope tube holds the
focal plane camera, which contains a single e2v CCD-22 detector.  A
thermal radiator mounted on the anti-solar side of the spacecraft is
coupled to a thermo-electric cooler (TEC) designed to cool the
detector to -100 C.  

Table~\ref{tbl:instrument_characteristics} 
gives the basic design parameters of the XRT.  The effective
area and sensitivity were verified by an end-to-end X-ray
calibration of the assembled instrument at the Panter X-ray
Calibration Facility in September 2002.  

\begin{table}
\caption{XRT Instrument Characteristics}
\label{tbl:instrument_characteristics}
\begin{tabular}{ll} \hline
Telescope: & Wolter I (3.5 m focal length) \\

Detector: & e2v CCD-22 \\
Detector Format: & 600 $\times$ 600 pixels \\
Pixel Size: & 40 $\mu$m $\times$ 40 $\mu$m \\
Readout Modes: & Image (IM) mode \\
   & Photodiode (PD) mode \\
   & Windowed Timing (WT) mode \\
   & Photon-Counting (PC) mode \\
Pixel Scale: & 2.36 arcseconds/pixel \\
Field of View: & 23.6 $\times$ 23.6 arcminutes \\
PSF: & 18 arcseconds HPD @ 1.5 keV \\
    & 22 arcseconds HPD @ 8.1 keV \\
Position Accuracy: & 3 arcseconds \\
Time Resolution: & 0.14 ms, 1.8 ms, or 2.5 s \\   
Energy Range: & 0.2 - 10 keV \\
Energy Resolution: & 140~eV @ 5.9 keV (at launch) \\
Effective Area: & $\sim 125$ cm$^2$ @ 1.5 keV \\
   & $\sim 20$ cm$^2$ @ 8.1 keV \\
Sensitivity: & $2 \times 10^{-14}$ \flux in 10$^4$ seconds \\
Operation: & Autonomous \\
\hline
\end{tabular}
\end{table}

The design and scientific capabilities of the XRT instrument are
summarized in the following sections.

\section{SCIENCE REQUIREMENTS}

There are three primary requirements that drive the design of the XRT:
rapid, accurate position determination; moderate resolution
spectroscopy; and lightcurves with high timing resolution.

\begin{figure}
\centering
\includegraphics[bb=17 354 385 692,clip=true,width=3in,angle=0]{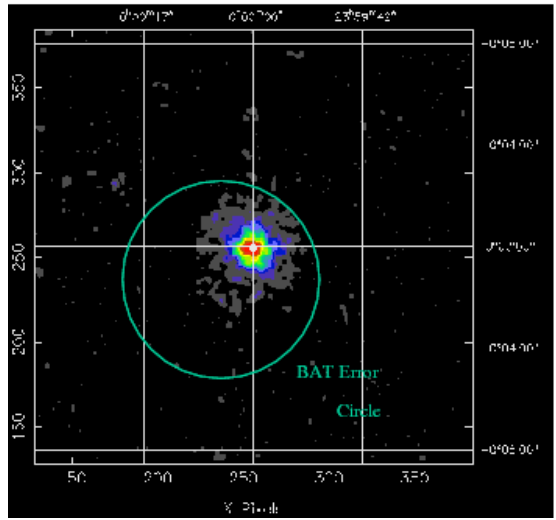}
\caption{Simulation of a typical XRT early-afterglow observation (about
  100s after the burst), using the measured PSF of the XRT mirrors.
  A 4 arcminute BAT error circle is overlaid for comparison.}
\label{fig:sim_image}
\end{figure}

\subparagraph{GRB Position Determination:}  The XRT is required to measure afterglow
positions with 
accuracy better than 5 arcseconds within 100 s of a
burst alert from the BAT instrument.
The spacecraft will slew to the BAT position in
20-75 s, depending on the position of the GRB on the sky.
Figure~\ref{fig:sim_image}
shows a simulated XRT image of a GRB, made using ray-tracing code
developed for JET-X and {\BeppoSAX} and incorporating the measured
PSF of the XRT mirrors.  The mirror PSF has a 15
arcsecond Half-Power Diameter (HPD) at the best on-axis focus (at 1.5
keV).  It is slightly defocused in the XRT in order to provide a more
uniform PSF over the entire field of view, and the instrument PSF is
18 arcseconds (HPD) on-axis at 1.5 keV \cite{9}.  The centroid of a point source image
can be determined to sub-arcsecond accuracy in detector coordinates,
given sufficient photons \cite{10}.  Based on {\BeppoSAX} and {\RXTE} observations
of X-ray counterparts of GRBs, we expect that most GRBs observed by
\Swift\ will have prompt X-ray fluxes of roughly 0.5--5 Crabs in the
0.2--10 keV band (Figure~\ref{fig:decay}).  XRT calibration data show that the XRT
will obtain source positions accurate to 1-3  arcseconds in detector
coordinates for typical afterglows within 5 seconds of target
acquisition.  When this position is referenced back to the sky, the
expected uncertainty is 3-5 arcseconds, due primarily to
the alignment uncertainty between the star tracker and the XRT.  In
order to minimize this error term, the star trackers are mounted on
the XRT forward telescope tube (Figures~\ref{fig:layout2},~\ref{fig:photo1}), and a Telescope
Alignment Monitor (TAM; \opencite{TAM}) measures flexing of the telescope tube with
subarcsecond accuracy.

\subparagraph{Spectroscopy:} X-ray spectroscopy can constrain
important properties of
the GRB/afterglow.  In the standard fireball model \cite{MR97}, the GRB
is produced by internal shocks in the relativistic fireball and the
afterglow is produced by external shocks with the ambient medium.
While most of the X-ray luminosity of the afterglows arises from
non-thermal synchrotron emission from the external shock, there have
been reports of X-ray line emission from GRB afterglows
(\opencite{18}; \opencite{19}). These
may be the result of thermal emission or of X-ray reflection (\opencite{L99}).
Observations of the X-ray spectrum may therefore detect emission
lines, which can provide direct information on such parameters as the
composition and ionization structure of the shocked gas, as well as
the redshift of the GRB.  Absorption edges from surrounding
unshocked gas will be observable over a wide range of column
densities (\opencite{L02}), and these edges can also provide both redshift and
abundance information.

\begin{figure}
\centering
\includegraphics[bb=50 0 574 720,clip=true,width=3in,angle=270]{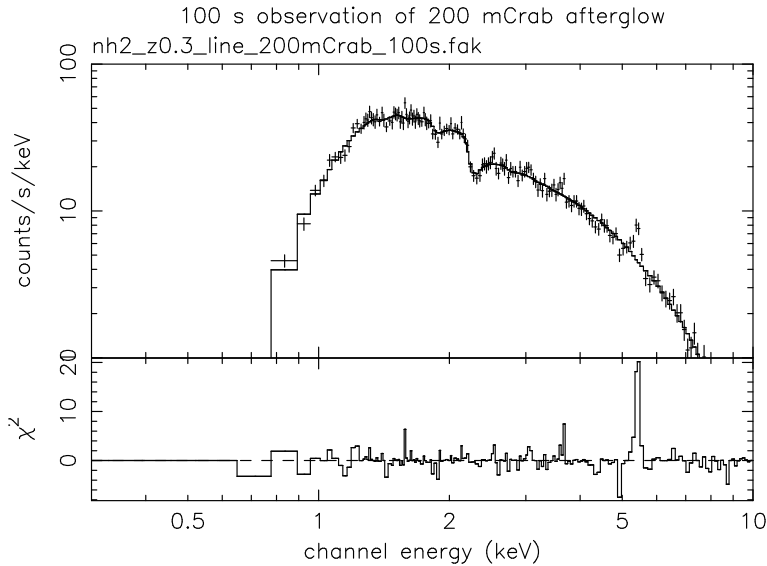}
\caption{Simulated spectrum from a 100 s XRT observation of a 200
  mCrab afterglow at z=0.3, assuming a power law spectrum plus a
  redshifted Gaussian Fe line ($E_{rest}=6.7$ keV).}
\label{fig:simulated_spectrum}
\end{figure}

The XRT energy resolution at launch was about 140~eV at 6~keV
(Figure~\ref{fig:resolution}), and spectra similar to that shown in 
Figure~\ref{fig:simulated_spectrum} will be
obtained routinely.  The strong Fe emission line in this simulated
spectrum ($E_{rest}=6.7$ keV) enables a measurement of the redshift that is
accurate to about 10\%.  
The XRT transmits unprocessed spectra to the
ground through the Tracking and Data Relay Satellite System (TDRSS) 
within 5--10 minutes of each BAT burst
alert to facilitate mission planning and follow-up observations of
afterglows with interesting X-ray spectra.  [These unprocessed spectra
have no event recognition applied and are
suitable only for qualitative purposes.  Fully processed quick-look
science data will be available from the
Swift Data Center several hours after each observation for use in
scientific analysis.]

The XRT readout modes are designed to allow spectroscopy
for sources up to $\sim 6 \times 10^{-8}$ \flux\ (0.2--10~keV).  
Sources brighter than this will be piled up
even in the fastest readout mode, but we are developing techniques for
analyzing these piled-up data (see \S \ref{sec:pileup}).  For more information on XRT readout
modes, see \inlinecite{11}.

\subparagraph{Light Curves:} The XRT is required to provide accurate
photometry and light curves with at least 10 ms time resolution. Two
timing modes have been implemented to meet this requirement: Photodiode (PD) mode and
Windowed-Timing (WT) mode.  

Photodiode mode is based on a timing mode
originally developed for the JET-X instrument.  It provides the best time
resolution (0.14 ms), but integrates the count rate over the entire
CCD and therefore provides no spatial information. 
It is suitable for use in uncrowded fields, or fields dominated by a
single bright source, and can measure
source fluxes as bright as 65 Crabs.  Because photodiode mode
combines data from the entire CCD, the on-board calibration sources
(located over the corners of the CCD; see \S~\ref{sec:fpca})
contaminate PD mode spectra, and it is important to subtract this
instrumental background accurately before
beginning astrophysical analysis.

Windowed timing (WT) mode is similar to the timing modes available on the
{\it Chandra}/ACIS instrument and the {\it XMM}/EPIC MOS cameras, and
provides about 2 ms time resolution with 1-D
spatial resolution within a window 8 arcminutes wide.  
A more detailed discussion of WT mode is presented in \S\ref{sec:modes}.

Flux accuracy for light curves has a systematic uncertainty of about 10\% for a wide range of
incident fluxes up to 
$8 \times 10^{-7}$ \flux.  Figure~\ref{fig:Panter_lightcurve} shows a light curve from Panter
calibration data with the incident flux ranging over five orders of
magnitude.  XRT light curves begin as soon as the GRB enters the XRT
field of view.

\section{INSTRUMENT DESCRIPTION}
\label{sec:design}

\subsection{Structure}
The layout of the XRT is shown in Figures 2 and 3.  The XRT structure
is designed around an aluminum Optical Bench Interface Flange (OBIF),
with a forward telescope tube supporting the star trackers and the XRT
Door Module and an aft telescope tube supporting the Focal Plane
Camera Assembly.  The total mass of the XRT, excluding the heat
pipe/radiator system, is 198.1 kg.

\subparagraph{Optical Bench Interface Flange:} The 
OBIF (Figure 11) is the primary structural element of the XRT and
is responsible for supporting the forward and aft telescope tubes, the
mirror module, the electron deflector, and the TAM optics and camera.
It also provides the mounting points to the \Swift\ Observatory.

\subparagraph{Telescope Tube:} This 508 mm diameter graphite fiber/cyanate ester
tube, manufactured by ATK, is composed of two sections.  The carbon
fiber layup is designed to minimize the longitudinal coefficient of
thermal expansion so that temperature gradients will not adversely
affect the alignment or focus.  The composite tube is lined internally
with an aluminum foil vapor barrier to guard against outgassing of
water vapor or epoxy contaminants into the telescope interior.  The
rear tube supports the Focal Plane Camera Assembly, and incorporates
internal optical baffles.  The forward telescope tube section encloses
the mirrors and supports the door assembly and the star trackers.
The internal volume of the telescope tube is vented to space through a
baffled vent in the Focal Plane Camera Assembly.

\subparagraph{Door:} The telescope tube is sealed at the forward end for ground
operations and launch by a single-shot
door assembly built by Starsys, which is designed to protect the
X-ray mirrors from contamination.
This door was opened about two weeks after launch, which gave the
spacecraft time to outgas before the XRT mirrors were exposed to the
ambient conditions surrounding the spacecraft.

\subsection{Optics}
\begin{figure}
\centering
\includegraphics[bb=133 25 500 763,clip=true,height=3in,angle=90]{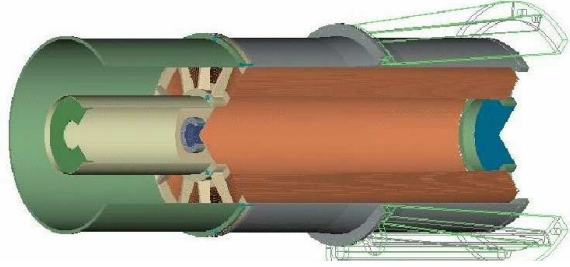}
\caption{Cutaway view of XRT mirror assembly.  The thermal baffle, on
  the left, consists of two cylinders with foil heaters mounted on
  them to control the mirror temperature by replacing the heat
  radiated to space.  The mirror module (Fig.~\ref{fig:mirrors}) is supported by the conical mirror
spacer (shown as line drawing for clarity), which attaches it to the optical bench interface flange.}
\label{fig:mirror_cutaway}
\end{figure}

\begin{figure}
\centering
\includegraphics[bb=46 8 559 783,clip=true,height=3in,angle=90]{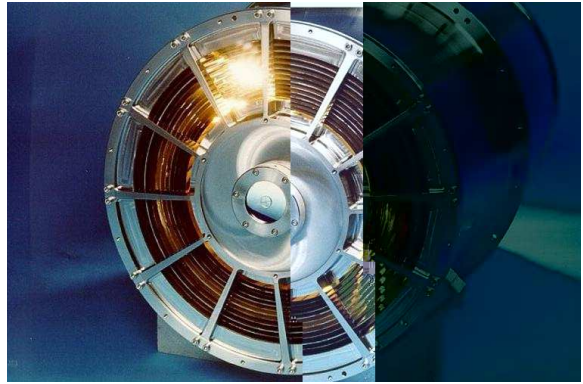}
\caption{XRT mirror module.  This is the FM3 mirror module originally built
  as a spare mirror set for the JET-X instrument on {\it Spectrum X-$\Gamma$}.  It
  consists of 12 nested Wolter-I grazing incidence mirrors held in
  place by front and rear spiders.}
\label{fig:mirrors}
\end{figure}

\subparagraph{Overall description:} The XRT mirror assembly is shown
in Figure~\ref{fig:mirror_cutaway}.  It consists of
the X-ray mirror module (Figure~\ref{fig:mirrors}), a thermal baffle mounted in front
of the mirrors (to the left in Figure~\ref{fig:mirror_cutaway}), 
a mirror spacer (shown as a line drawing for clarity) that mates to the XRT optical bench
interface flange, and an electron deflector that mounts behind the
mirrors (not visible in this figure).

\begin{figure}
\centering
\includegraphics[bb=29 178 580 660,clip=true,height=3in,angle=0]{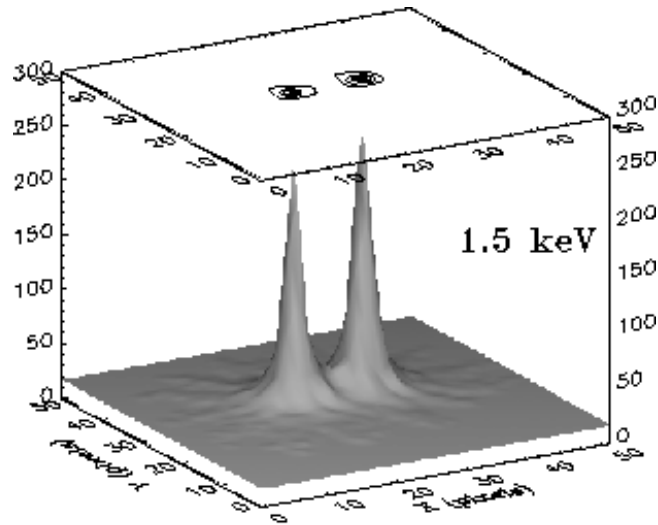}
\caption{
Image of two point sources displaced by 20
  arcseconds, made during mirror calibration at the Panter calibration facility.  Although the
  mirror HPD is 18 arcseconds, the PSF is sharply peaked with a FWHM of
  about 7 arcseconds.  The horizontal axes are in pixels and the
  vertical axis is in arbitrary units.}
\label{fig:psf}
\end{figure}

\subparagraph{XRT Mirrors:} The XRT uses the FM3 mirror set
(Figure~\ref{fig:mirrors}) built and
calibrated for the JET-X program (\opencite{Citterio96}; \opencite{8}).  
The mirror module was developed at the Brera Observatory and was
manufactured by Medialario.
It has
12 concentric gold-coated electroformed Ni shells with focal length
3500 mm.  The shells are 600 mm long with diameters ranging from 191
to 300 mm.  The effective area and point spread function of the
mirrors have been measured for a variety of energies and off-axis
angles, and were recalibrated at the Panter facility in July 2000.
A calibration image of two sources displaced by 20 arcseconds (Figure~\ref{fig:psf})
shows the very sharply peaked PSF.  As noted above, the detector
is slightly defocused in the flight instrument in order to optimize
the angular resolution over the field of view.  Results of the
end-to-end calibration of the XRT effective area and PSF are given in
\inlinecite{13} and \inlinecite{9}, respectively.

\subparagraph{Thermal Baffle:} A thermal baffle in front of the mirror
prevents temperature gradients in the mirror that can distort its
figure and degrade its PSF. This system is
discussed in more detail in \S\ref{sec:thermal_design}.

\begin{figure}[bth]
\centering
\includegraphics[width=2.5in,angle=270]{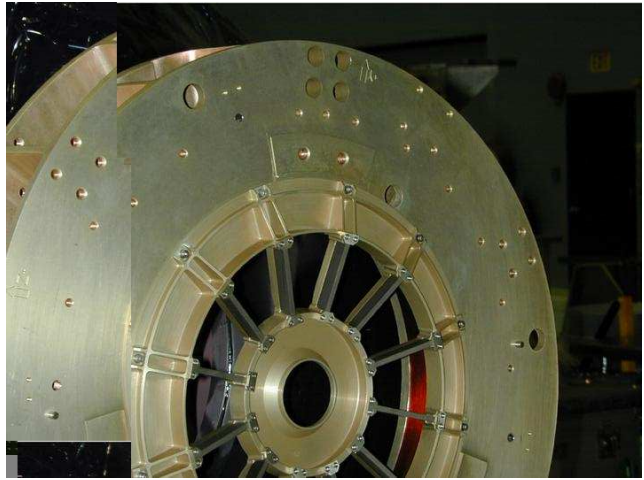}
\caption{The XRT electron deflector, mounted on the rear face of the
  OBIF.  The electron deflector consists of 12 rare earth bar magnets
  mounted on radial spokes
  aligned with the mirror spiders.  These provide a high azimuthal
  magnetic field across the optical path with very low dipole moment.}
\label{fig:electron_deflector}
\end{figure}

\subparagraph{Electron Deflection Magnets:} The electron flux in our
$584 \times 601$~km, 20$\degdot$6 inclination 
orbit will produce a variable detector background. An electron
deflector, consisting of a system of 12 rare earth magnets, is
installed on the OBIF (Figure~\ref{fig:electron_deflector}) just behind the rear face of the
mirror module to prevent electrons that pass through the mirror from
reaching the detector.  The design is scaled down from the \XMM\
electron deflectors, and has a cutoff energy of ~22 keV with near-zero
dipole moment.

\subsection{Focal Plane Camera Assembly (FPCA)}
\label{sec:fpca}
\begin{figure}
\centering
\includegraphics[bb=115 249 750 725,clip=true,width=\maxfloatwidth, angle=0]{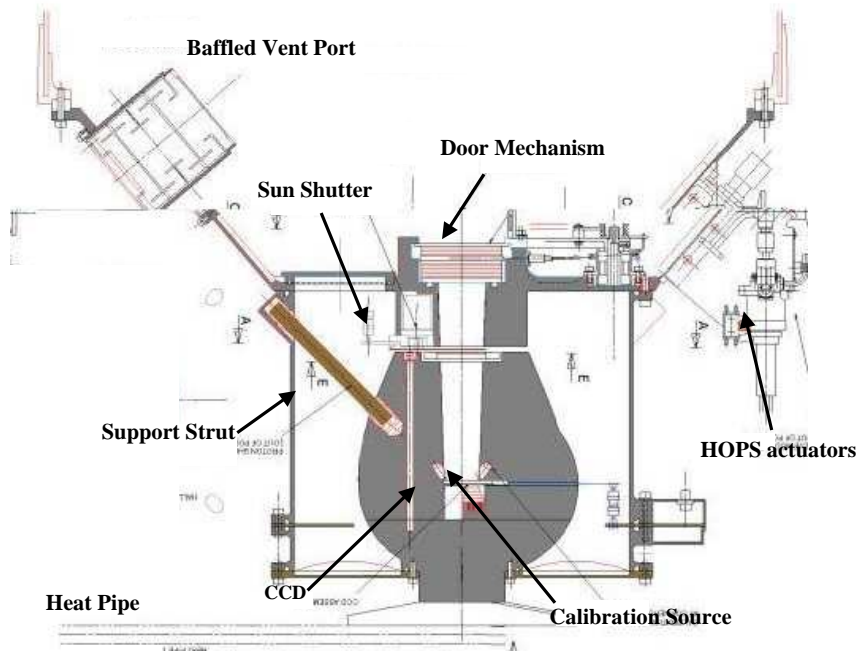}
\caption{Focal Plane Camera Assembly cross-section.  The entire proton
  shield is cooled to a temperature of T$<$-50 C by the
  HRS (a radiator/heat pipe system).  A TEC integrated into the CCD flight
  package is designed to maintain the CCD temperature at -100 C, using a closed-loop
  cooling control system.  The FPCA also houses the XRT baffled vent.
An autonomous sun shutter protects the detector and filter from
accidental exposure to the sun in the event of an attitude control
failure.
The optical light blocking filter sits immediately below the sun shutter.}
\label{fig:fpca_drawing}
\end{figure}

\begin{figure}
\centering
\includegraphics[width=2.5in,angle=270]{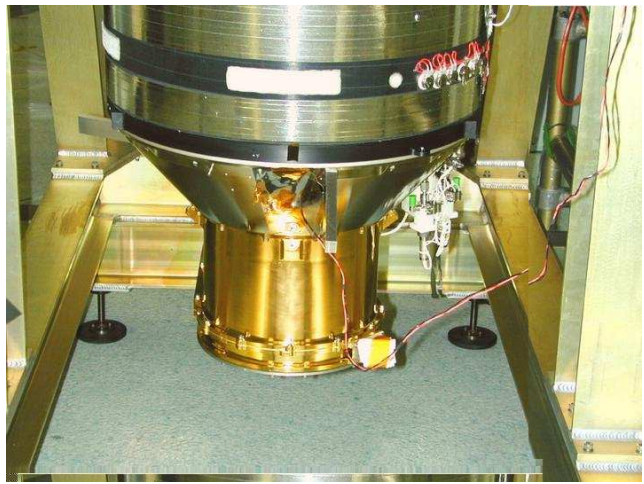}
\caption{The XRT Focal Plane Camera Assembly, mounted on the
  bottom of the XRT instrument.}
\label{fig:fpca}
\end{figure}

The FPCA (Figures~\ref{fig:fpca_drawing} and \ref{fig:fpca}) provides a vacuum enclosure for the CCD and
optical blocking filter during ground operations and launch, radiation
shielding
for the CCD against trapped protons and cosmic rays, and cooling for
the detector. The cryostat has a single-shot door mechanism utilizing
redundant Starsys actuators. 

The cryostat is attached to a conical interface section that mounts
onto the rear tube, supports the cryostat door control hardware and
radiator interface, and incorporates a baffled venting system.  The
vent port allows the telescope internal volume to vent during vacuum
testing and launch, while preventing scattered light from entering the
CCD enclosure. 

A novel feature of the FPCA is the sun shutter, which is a safety
mechanism designed to autonomously protect the CCD and filter from
accidental solar illumination in the event of an attitude control
failure.  The sun shutter is powered by a Ga-As array mounted at the
top of the inner thermal baffle, which provides power to the sun shutter
electronics and closes the shutter if the spacecraft slews to within
$\sim 30$ degrees of the Sun,
even if the instrument is turned off.  
The sun shutter can also be opened or closed by telecommand, but it
will never be intentionally closed after launch.

\begin{figure}
\centering
\includegraphics[width=2.75in,angle=270]{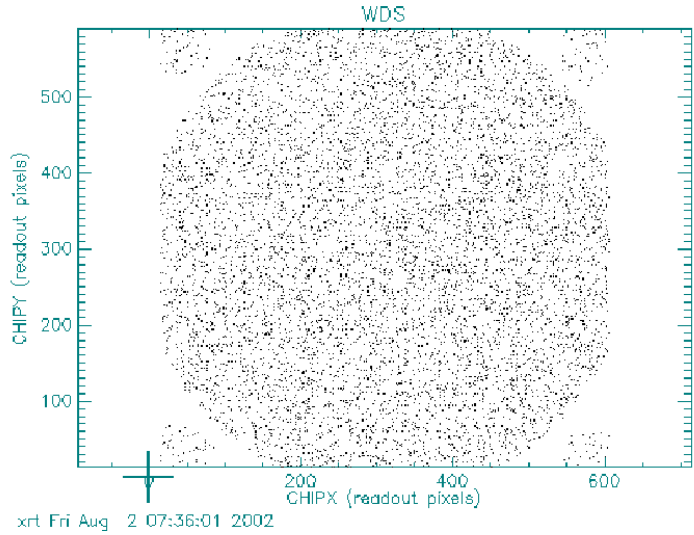}
\caption{XRT image of the on-board calibration sources.  Each spot
  represents a single X-ray photon.  The large central circle shows
  the instrument field of view illuminated by an $^{55}$Fe calibration
  source mounted on the inside of the camera door.  The four circles of 
X-rays in the CCD corners are from sources that illuminate the CCD  
corners continuously during flight.}
\label{fig:cal_image}
\end{figure}

\begin{figure}
\centering
\includegraphics[bb=89 62 523 712,clip=true,width=2.75in,angle=90]{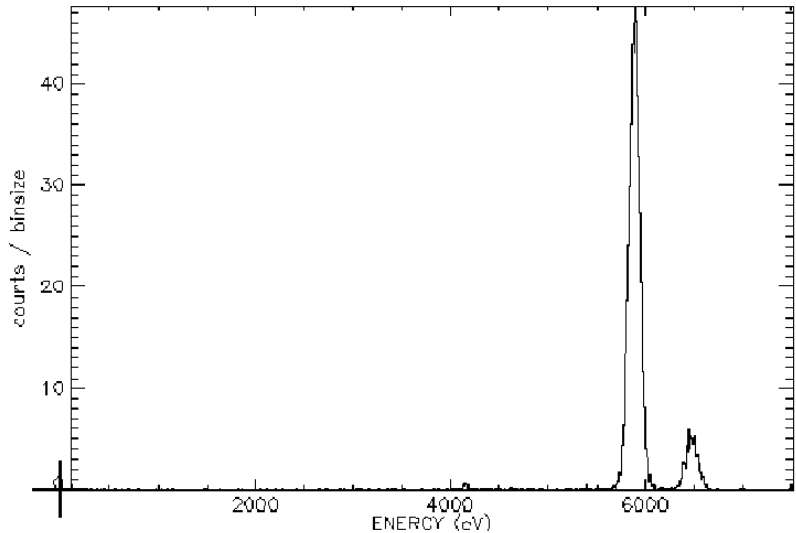}
\caption{XRT spectrum of Mn K$\alpha$ and K$\beta$ X-rays from the
  door source (PC mode).  The FWHM of the 5.9 keV line is 135--140~eV, depending
  on event grade.}
\label{fig:spectrum}
\end{figure}

The FPCA also houses two sets of $^{55}$Fe calibration sources, which
undergo inverse $\beta$-decay and produce Mn K$\alpha$ and K$\beta$ X-rays at 5.9 and 6.5 keV.  An image
collected during instrument thermal vacuum testing in July 2002 is shown in
Figure~\ref{fig:cal_image}.  The central circular region shows the field of
view (FOV) of the instrument (defined by the circular optical blocking
filter housing, but slightly truncated by the edges of the CCD), which
was illuminated by a bright $^{55}$Fe calibration source located on
the inside of the camera door during ground testing and the first
three weeks of flight operations.  
The spectrum of these
events is shown in Figure~\ref{fig:spectrum}.  
This source was moved out
of the FOV when the camera door was opened about three weeks after
launch.
The four circular clusters of X-ray events in the
corners of the detector are the in-flight calibration sources, which
permit us to measure resolution, gain, and charge transfer efficiency
throughout the mission.


\subsubsection{Optical Blocking Filter}  
\begin{figure}
\centering
\includegraphics[width=2.5in,angle=270]{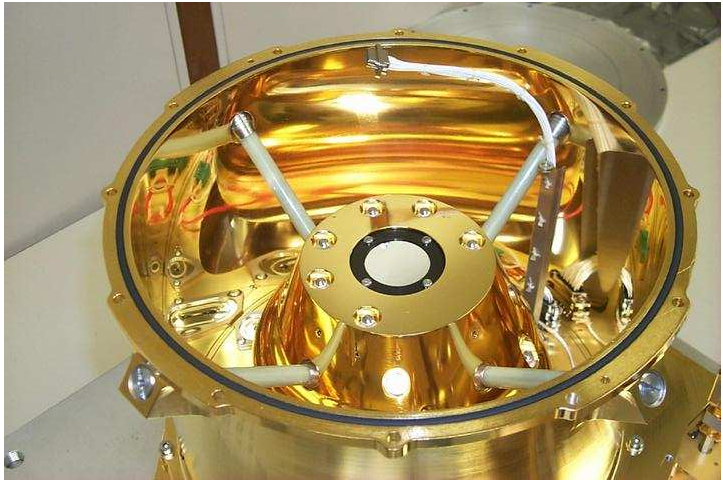}
\caption{Interior of the FPCA, showing the lower half of the camera.  
The circular UV/optical blocking filter in the center of the photo is mounted at the top of the 
lower proton shield, which is mechanically supported by four
fiberglass struts.  
The black ring around the filter is the filter frame, which defines 
the instrument field of view.  
The sun shutter assembly (not shown) mounts immediately above the filter.}
\label{fig:fpca_interior}
\end{figure}

A thin Luxel filter is installed
in front of the CCD to block optical light.  The filter is similar to
those used on the {\it Chandra}/ACIS and \XMM/EPIC instruments.  It consists
of a single fixed polyimide film 1840\AA\ thick, coated on one side with
488\AA\ 
of aluminum.  The optical transmission of the filter is about
$2.5 \times 10^{-3}$.  The filter can be seen in the center of Figure~\ref{fig:fpca_interior}, which
shows the interior of the FPCA.
The relatively high filter transmission allows optical light from
bright stars to contaminate some fields, and also allows optical light
scattered from the bright Earth into the camera when viewing near the
Earth limb in daylight.  Data cleaning can eliminate both sources of
contamination for most directions on the sky.

\subsubsection{XRT Detectors}
\subparagraph{CCD Architecture:} The CCD-22 detector, designed for the
EPIC MOS cameras on {\it XMM-Newton} by e2v (formerly known as EEV), is a
three-phase frame-transfer device, which utilizes high resistivity
silicon and an open-electrode structure \cite{3} to achieve a useful bandpass
of 0.2 to 10 keV \cite{4}. It
has an imaging area of ~2.4 x 2.4 cm. The image section of each CCD is
a $600 \times 600$ array of $40 \mu$m $\times 40 \mu$m pixels, each pixel corresponding to
2.36 arcseconds in the \Swift\ focal plane. The storage region is a 
$600 \times 600$ pixel array of $39 \mu$m $\times 12 \mu$m pitch.  

\begin{figure}
\centering
\includegraphics[width=3in,angle=270]{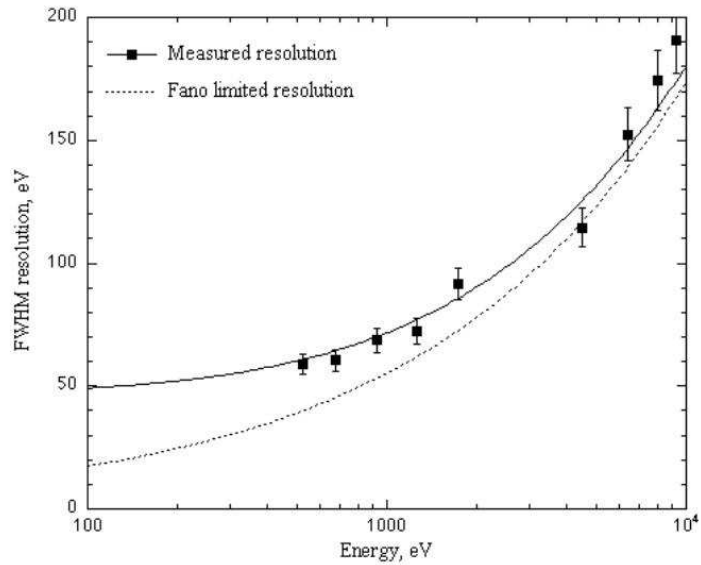}
\caption{CCD-22 energy resolution above 500 eV.}
\label{fig:resolution}
\end{figure}

\subparagraph{Energy Resolution:} The energy resolution of the CCD-22 is
shown in Figure~\ref{fig:resolution}, in which the dotted line is the
ideal
(Fano-limited) resolution, the solid line is the predicted resolution
for an EPIC MOS CCD and the points are measurements from a typical
flight device. Below $\sim 500$~eV the effects of charge trapping and loss
to surface states become significant.

\subparagraph{Quantum Efficiency:} The QE of an X-ray CCD is determined at low energies
by the gate structure and at high energies by the depletion depth.  A
special ``open-gate'' electrode structure gives the CCD-22 excellent low
energy quantum efficiency, while high resistivity silicon provides a
depletion depth of 30 to 35$\mu$m to give good QE at high energies.
Quantum efficiency measurements of these detectors have been made at
the University of Leicester and at the Orsay Synchrotron. These
measurements have been used to validate our Monte Carlo model of the
CCD-22 spectral response.  The model was used to
construct XRT response matrices for all readout modes and for various
selections of event grades \cite{14}.

\subparagraph{Radiation Hardness:} CCDs are susceptible to both ionizing
and non-ionizing radiation damage, but the primary area of concern for
an X-ray photon-counting CCD spectrometer in low Earth orbit is
displacement damage caused by proton irradiation, which results in the
creation of electron traps in the silicon lattice.  These traps
degrade the charge transfer efficiency of the device over time and
directly degrade its energy resolution.  We note that the ability of
the CCD to measure accurate source positions and lightcurves is
unaffected by this proton damage until doses far in excess of the
total \Swift\ mission dose are reached.

We performed a radiation study for the XRT in the baseline \Swift\
orbit of 600 km altitude and 22$\degrees$ inclination.  This study indicated
that the equivalent 10 MeV proton flux seen by the XRT detectors will be
$3.4 \times 10^8$ protons cm$^{-2}$ yr$^{-1}$.  At the actual orbital
inclination of 20$\degdot$6 the dose will be somewhat less.
The expected total dose for the XRT in a
nominal two-year mission is therefore comparable to the total mission fluence of
$5 \times 10^8$ protons cm$^{-2}$ expected for \XMM.  Considerable analysis and
laboratory test data from the EPIC program are therefore directly
applicable to the XRT.  Laboratory measurements using CCD-22s
irradiated to $2.5 \times 10^8$ 10 MeV protons cm$^{-2}$ have been used to verify that
the XRT will have energy resolution of better than 300 eV 
at 6 keV after 3 years on-orbit \cite{radiation_damage_report}.   

\subsection{Thermal Design}
\label{sec:thermal_design}
\subparagraph{Cryostat:} The instrument was designed to operate the X-ray CCD detectors at
-100 C ensure low dark current and to reduce sensitivity to radiation
damage. 
A failure of the TEC Power Supply shortly after launch prevents us from
reaching this temperature or from maintaining a stable temperature
on-orbit.  In practice the CCD is now being cooled passively through the HRS 
to temperatures ranging from about -50C to about -70C,
depending on the observatory orientation.

\subparagraph{Heat Rejection System (HRS):} The HRS consists of
redundant ethane heat pipes coupled to a 900 cm$^2$ radiator coated with
AZW-LA-II paint.  The radiator is carefully configured and integrated
with the external spacecraft design to achieve the required low
temperature of ${\rm T} < -50\degrees$~C in low Earth orbit (depending on orbital
parameters and spacecraft orientation).  The HRS is 
connected to the cold finger of the FPCA cryostat, where it
provides cooling to the TEC.  Although the XRT was not designed for
solely passive cooling, the on-orbit performance being achieved meets
the design requirements.

\subparagraph{Telescope thermal design:} The XRT uses both passive
design features and active thermal control to achieve the high degree
of dimensional stability required for arcsecond alignment
tolerances. Operational heaters on the telescope tube are
controlled by software thermostats in 36 independent zones (arranged in 9
bands, each with four quadrants) to maintain a constant longitudinal
thermal gradient and less than 2~C azimuthal thermal gradient on the
telescope tube to prevent bending of the tube.   Within each quadrant a
temperature sensor measures the temperature of that quadrant.  These data are
monitored by our housekeeping circuits once per second.  Every 30
seconds the XRT software compares each quadrant's temperature to an
adjustable setpoint for that quadrant and turns the corresponding
heater on or off accordingly.  The setpoint and deadband (hysteresis)
of each quadrant are independently adjustable.  This system performed
well during thermal balance tests.

Survival heaters are designed to prevent damage to the instrument when
the instrument power is off.  Survival heaters are located at critical
points on the telescope tube and on the electronics box.

\subparagraph{Mirror thermal design:} The mirror module must be
maintained at 18~C~$\pm$~0.5~C with a delta of less than 1 C from front
to back in order to prevent degradation of the mirror PSF.  The
mirrors are not heated directly; instead, an actively controlled
thermal baffle in front of the mirrors replaces the heat radiated to
space by the mirrors.  
The baffle consists of an inner cylinder and an outer cylinder with
heaters, thermostats, and temperature sensors.
The temperature profile of the baffle
is actively controlled at two longitudinal locations by pairs of operational
control heaters with software thermostats, which balance the heat lost to
space by the mirrors.  
Thermal control is provided by the flight software
using adjustable setpoints in the same way that the operational tube
heaters are controlled. During thermal balance tests the flight
mirrors were replaced with an identical mirror set with thermal
instrumentation, and the mirror
baffles successfully maintained the thermal gradient along the mirrors
at a few tenths of a degree.

When the instrument is turned off the mirror baffle
temperature is maintained by redundant survival heaters controlled by
mechanical thermostats.  This prevents the mirrors from becoming too
cold in the event of an observatory safehold.

\subsection{Telescope Alignment Monitor (TAM)}
\label{sec:TAM}
Although we expect the composite tube/heater thermal system to provide
a stable platform, the XRT will be operating in an unusually dynamic
thermal environment, due to the rapid slewing of the observatory.  
As an additional measure to correct for any
thermal variations affecting the instrument alignment, the XRT has an
internal Telescope Alignment Monitor (TAM) designed to actively
measure any remaining alignment errors between the XRT boresight and
the star trackers.  

\begin{figure}
\centering
\includegraphics[bb=115 83 485 702,clip=true,width=2.75in,angle=270]{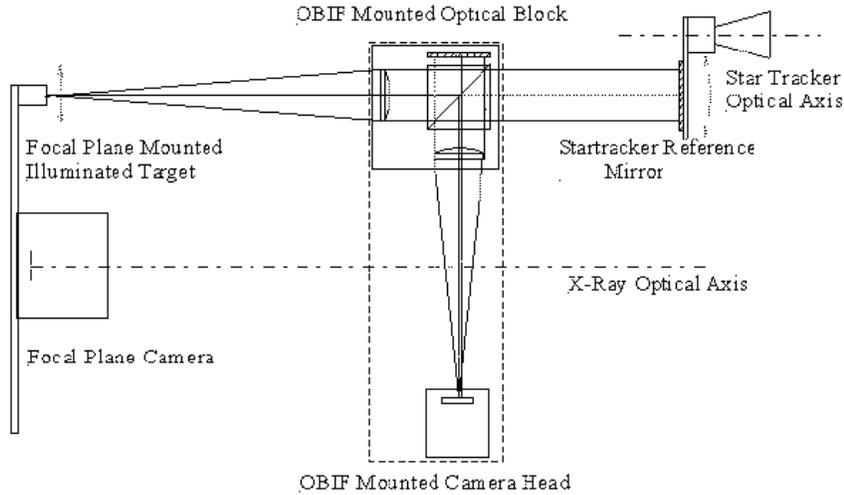}
\caption{Optical design of the Telescope Alignment Monitor.}
\label{fig:tam}
\end{figure}

The TAM (Figure~\ref{fig:tam}) uses one of two redundant LEDs mounted on the FPCA, which
are observed by a small radiation-hard programmable CMOS active pixel
sensor camera developed for space applications by SIRA
Electro-Optics. The TAM camera and optics block are mounted on the
optical bench interface flange (OBIF), which is the hard mount point
for the forward and aft telescope tubes and the mirror module, and
which is mounted in turn to the \Swift\ optical bench.  The TAM camera
monitors two light paths: the direct path from the FPCA LED, and a
secondary path that includes a reflection from a mirror mounted to the
star tracker platform.  The direct path monitors lateral motion of the
FPCA relative to the OBIF, which defines (to first order) the XRT
boresight.  The secondary path monitors changes in the tilt of the
star trackers relative to the OBIF, and is sensitive to deviations in
the star tracker boresight.  These two paths produce two images of the
LED every five minutes, which are independently monitored and telemetered.  
The TAM centroids are calculated 
on-board and are used to calculate GRB positions to be telemetered to
the Gamma-ray Coordinate Network (GCN).
They are also recorded in the science telemetry stream for use
by ground processing software to construct accurate sky maps for
the XRT.

We have performed several tests to verify the accuracy and utility of
the TAM data.  We have measured the displacement of TAM centroid
positions as the telescope tube heaters were
used to bend the rear tube slightly by heating it asymmetrically.
Movement of the TAM image by less than 0.05 TAM pixels, 
corresponding to movement of the XRT boresight by
less than 0.5 arcseconds, is easily measurable. 

\subsection{Electronics Design}
\begin{figure}
\centering
\includegraphics[width=3.5in,angle=270]{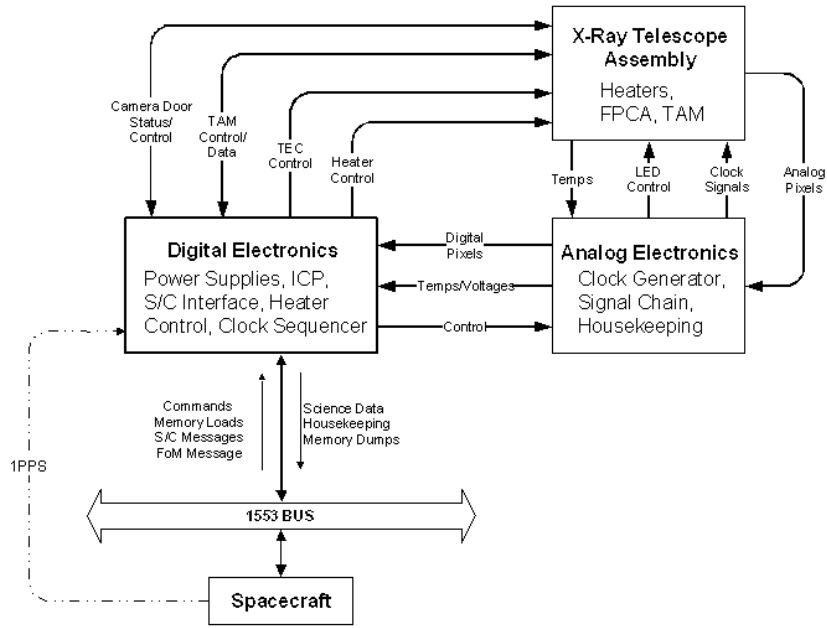}
\caption{Simplified XRT Electronics Block Diagram.}
\label{fig:electronics}
\end{figure}

The overall electronics design of the XRT is shown in the block diagram in 
Figure~\ref{fig:electronics}.  The electronics design has been discussed in some detail
previously (\opencite{15}; \opencite{16}). The instrument is controlled by a RAD6000 board
designed originally for the Mars98 mission.  The CCD is controlled by
a radiation-hard digital signal processor using software-controlled
waveforms, allowing us to implement a very flexible instrument with
readout modes optimized to deal with sources ranging over a very wide
dynamic range.  Dual signal
chains extract the CCD signal packets from the detector video signal,
using a correlated double sampler.  

\subsection{Automated Operations}
\label{sec:modes}
\begin{figure}
\centering
\includegraphics[width=3.25in,angle=0]{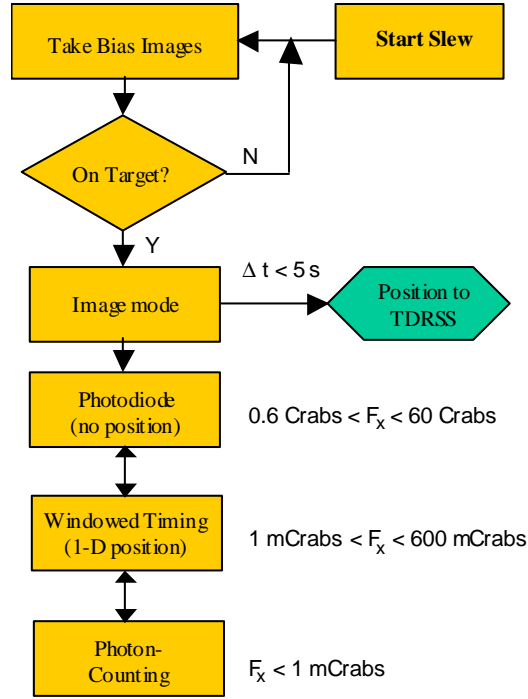}
\caption{Simplified XRT automation flow chart for a new GRB.}
\label{fig:flowchart}
\end{figure}

Because the \Swift\ spacecraft must respond rapidly to new targets, the
\Swift\ instruments must be able to operate autonomously.  This
poses interesting new problems for the instrument design.  Over the
course of a typical GRB observation, the burst/afterglow flux will
decrease by many orders of magnitude, and the XRT must be able to observe
over this wide dynamic range without detector saturation at early
times when the burst and afterglow are bright, while preserving as
much information about the incident photons as possible during sensitive observations of
faint afterglows at later times.
The XRT is
designed for completely autonomous operation, switching between
different readout modes according to the instantaneous count rate in
each CCD frame, as shown in Figure~\ref{fig:flowchart}.  
This algorithm was successfully tested at the
Panter X-ray facility using a variable source flux ranging over five
orders of magnitude, designed to mimic
the behavior we expect from real GRBs
(Figure~\ref{fig:Panter_lightcurve}).
The autonomous operation of
the XRT and its readout modes are described in more detail in
\inlinecite{11}.  
We briefly describe the readout modes here:

\begin{itemize}
\item[Image (IM) mode:] the CCD is operated like an optical
  CCD, collecting the accumulated charge from the target and reading
  it out without any X-ray event recognition.  For a typical GRB, this
  image will be highly piled up and will therefore produce no
  spectroscopic data, but it will produce an accurate position and a
  good flux estimate.  Image mode is operated with low gain to allow
  observations up to the full-well capacity of the CCD (in normal
  gain we are limited by the analog-to-digital converter range).
  Image mode uses either 0.1 or 2.5 second exposures, depending on the source
  flux, with the selection made automatically on-board.  
  Image mode can be used to determine on-board centroids for
  source  fluxes between  25 mCrabs and at least 45 Crabs.  The following data
  products are produced: the GRB centroid position and X-ray flux
  estimate, telemetered through TDRSS and distributed immediately to
  the community through the Gamma-ray burst Coordinate Network
  (GCN; \opencite{17}); a postage-stamp image ($2'\times 2'$), also telemetered via TDRSS
  and distributed through the GCN; and a compressed image (pixels
  above a threshold) in the normal science telemetry stream.  The
  X-ray flux estimate assumes a Crab-like spectrum.
\begin{figure}
\centering
\includegraphics[bb=0 317 800 740,clip=true,width=\maxfloatwidth,angle=0]{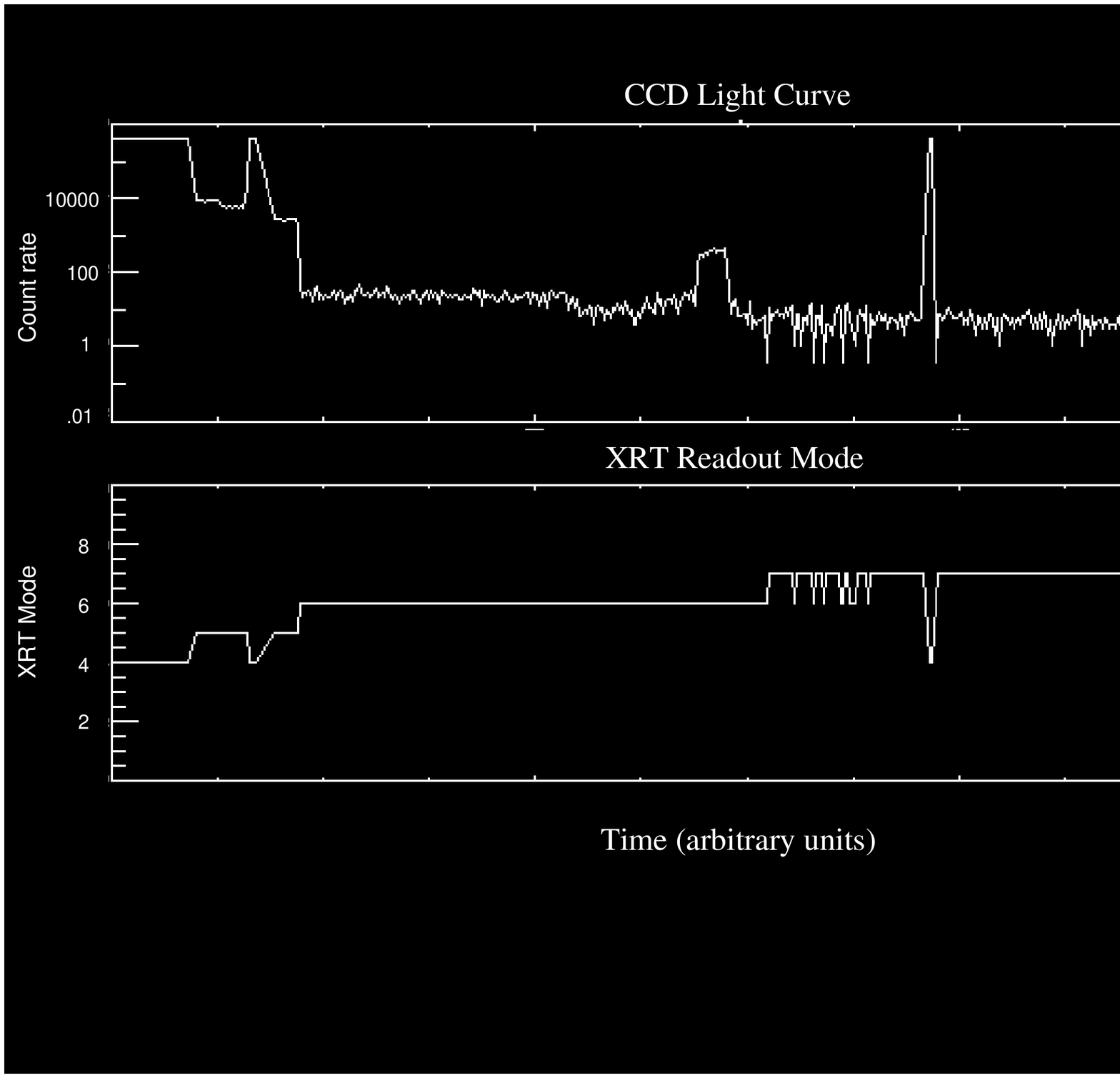}
\caption{XRT light curve during Panter tests, with incident flux
  ranging over 5 orders of magnitude.  The upper panel shows the XRT
  lightcurve; the vertical axis is the
  count rate (log scale), and the horizontal axis is CCD frame
  number.  The lower panel shows the XRT readout mode, which changes
  automatically in response to changes in the countrate. A value of 4
  represents PD mode (high rate), 5=PD mode (low rate), 6=WT mode, and 7=PC mode.}
\label{fig:Panter_lightcurve}
\end{figure}

\begin{figure}
\centering
\includegraphics[width=4.0in,angle=90]{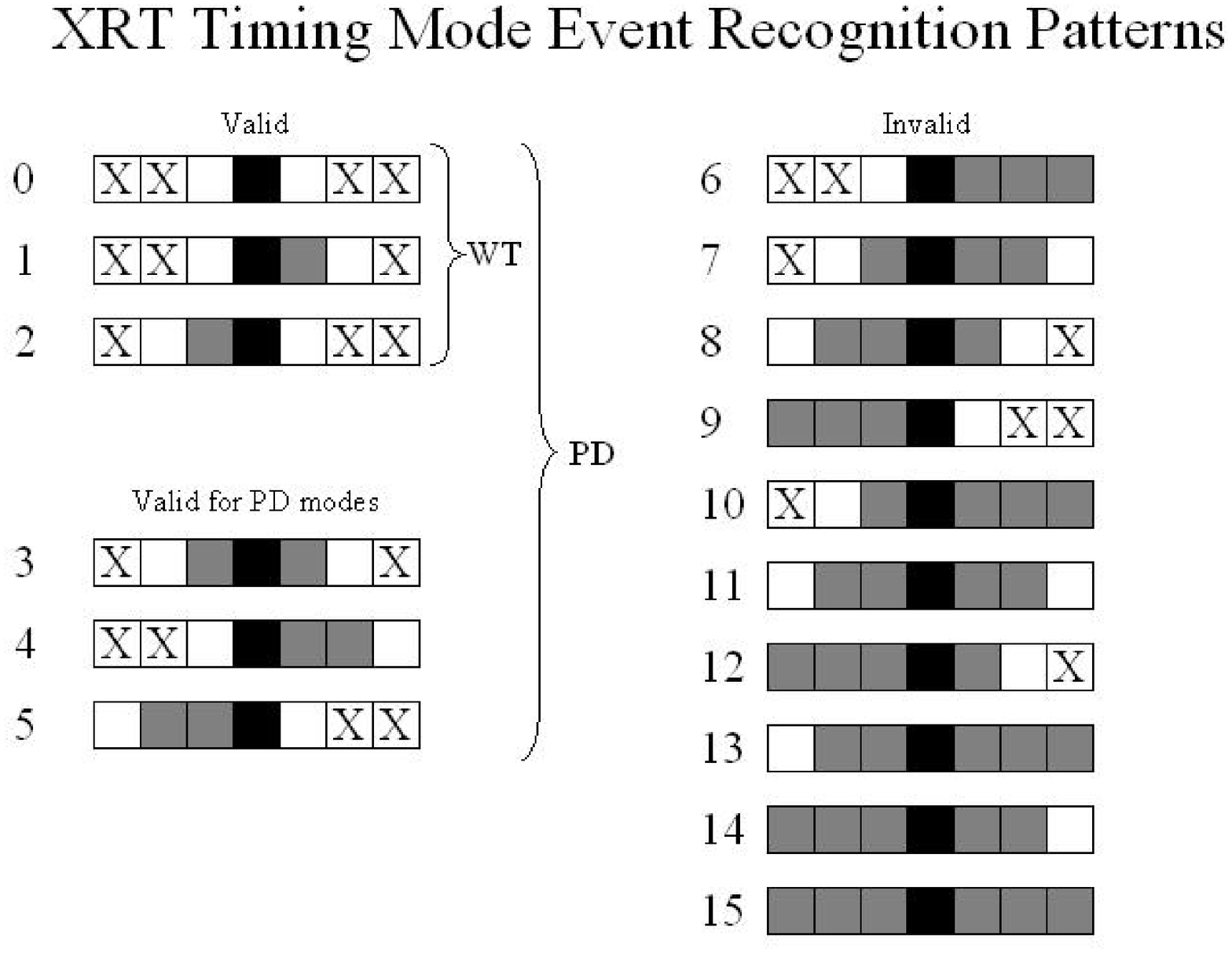}
\caption{XRT timing mode event grades.  For WT mode, only event grades 0-2
  represent valid X-ray events.  Valid events in PD mode include
  grades 0-5.  In both cases the black pixel represents the local
  maximum (the pixel containing the most charge), the gray pixels are
  those with signal levels above the event threshold, and the white
  pixels are those with signal levels below the event threshold.
  Pixels containing `X' are those for which the signal level does not matter.}
\label{fig:timing_grades}
\end{figure}

\item[Photodiode (PD) mode:] a fast timing mode designed to produce
      accurate timing information for extremely bright sources.  This
      mode alternately clocks the parallel and serial clocks by one
      pixel each.  Charge is accumulated in the serial register during
      each parallel transfer, with the result that each digitized
      pixel contains charge integrated from the entire field of view
      (although not simultaneously).  For the GRB case, where we
      expect the image to be dominated by a single bright source,
      photodiode mode produces a high-speed light curve with time
      resolution of about 0.14 ms.  This mode is useful for incident
      fluxes up to 60 Crabs, and has manageable pileup for fluxes
      below 2 Crabs.  Ground-processed data products in photodiode mode are FITS binary
      table files with the time, energy, and grade (or pattern) of
      each recorded event (unless the data are too piled-up to
      identify individual photon events).  Event grades for PD mode
      are described in Figure~\ref{fig:timing_grades}.

\item[Windowed Timing (WT) mode:] uses a 200 column window
      covering the central 8 arcminutes of the FOV (other window sizes
      are also possible).  Imaging
      information is preserved in one dimension, but the columns are
      clocked continuously to provide timing information in the
      trailed image along each column, at the expense of imaging
      information in this dimension.  Pixels are binned by 10$\times$
      along columns.  This mode has 1.8 ms time resolution for a 200
      column window.  It is useful for fluxes below 5000 mCrabs, and
      has minimal pileup below 1000 mCrabs.  Ground-processed data products are FITS
      binary table events files with the 1-D position, arrival time, energy,
      and pattern of each X-ray event.  Event grades for WT mode are
      described in Figure~\ref{fig:timing_grades}.

\item[Photon-counting (PC) mode:] retains full imaging and spectroscopic
      resolution, but time resolution is only 2.5 seconds.  PC mode
      uses a ``normal'' CCD readout sequence, in which the entire CCD
      is read out every 2.5 seconds, and processed on-board by
      subtracting a bias map and searching for X-ray events in $5
      \times 5$ pixel ``neighborhoods'' around each local maximum pixel.  It is
      useful for fluxes below 1 mCrab.  Ground-processed data products are FITS binary table
      events files
      with the 2-D position, arrival time, energy, $3 \times 3$ pixel
      neighborhood centered on the event, and the grade (or pattern)
      of each event.  Grades are recorded using a scheme similar to
      the \XMM/EPIC MOS pattern library.  Event grades for PC mode are
      described in Figure~\ref{fig:PC_grades} and are assigned during
      ground processing.
\end{itemize}

\begin{figure}
\centering
\includegraphics[height=4.0in,angle=90]{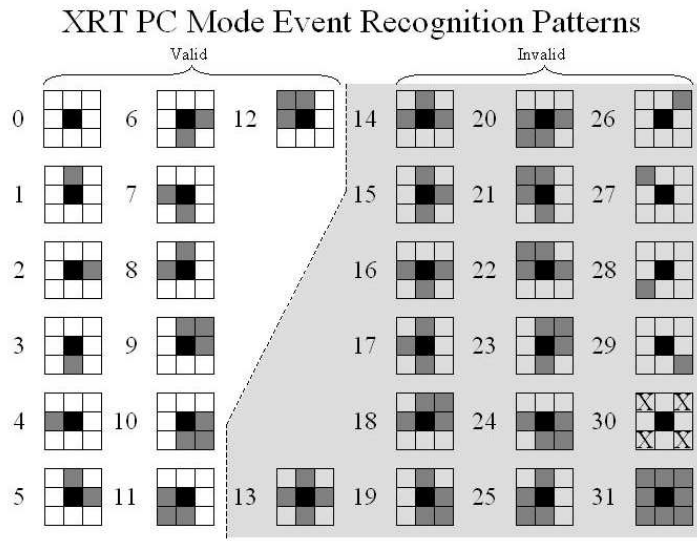}
\caption{XRT Photon-Counting mode event grades.  Grades 0-12 represent
valid X-ray events, while grades 13-32 represent primarily
charged particle background events or piled up events. For each grade, the black pixel represents the local
  maximum (the pixel containing the most charge), the gray pixels are
  those with signal levels above the event threshold, and the white
  pixels are those with signal levels below the event threshold.
  Pixels containing `X' are those for which the signal level does not
  matter.  Grade 32 (not shown here) is a catch-all for all patterns not specifically defined here.}
\label{fig:PC_grades}
\end{figure}

\subsection{XRT  Software}
\subsubsection{Flight Software}
There are two main pieces of flight software: 
\begin{itemize}
\item The Instrument Control Processor (ICP; this is the RAD6000 CPU)
      software is responsible for instrument control, data collection
      and processing, and spacecraft interfacing.  The ICP software is
      written in C and runs under the VxWorks operating system.  On
      boot-up, the code performs self-tests and enters Manual State
      with the CCD in Null mode (clocks running but no digitization of
      the data).  Manual state is used for engineering tests,
      calibration, or instrument configuration.  Once commanded into
      Auto State, the software automatically collects and processes
      data as described in more detail by \inlinecite{11}.
      Background processes ingest data from the TAM, perform memory scrubbing, and transmit
      telemetry data to the spacecraft.
\item The Clock Sequencer board produces the CCD clock waveforms using
      a Motorola 21020 Digital Signal Processor (DSP).  Clock
      waveforms are generated by an IDL program that
      uses a graphical user interface to produce 21020 DSP assembler
      code \cite{16}.  The code is then assembled and uploaded to the
      instrument.  On-board waveform programs support a number of
      readout modes, including diagnostic modes that
      facilitate checkout of the camera during ground tests and on
      orbit.
\end{itemize}

\subsubsection{Ground Processing and Data Analysis Software}

The Swift Data Center (SDC) at NASA/GSFC is responsible for the Level
0 software, which produces time-ordered telemetry data, and for the  Level 1
software, which produces standard format FITS event files.  

The Level 2/3 software is called the XRT Data Analysis Software
(\XRTDAS).
It consists of a set of
FTOOLS specifically developed for the XRT instrument by the ASI
Science Data Center (ASDC, Frascati, Italy) in collaboration with the
HEASARC at NASA/GSFC.  
{\XRTDAS} processes the FITS-formatted Level 1 XRT telemetry
data and 
generates higher-level scientific data products, including cleaned and
calibrated event files, images, spectra and light curves.  
The {\XRTDAS} input and output files are in FITS format and fully comply
with the NASA/GSFC Office of Guest Investigator Programs FITS standard
conventions.  

{\XRTDAS} is written using the C, Fortran and Perl languages.
It uses the HEASARC calibration database (CALDB) mechanism, and runs on most popular Unix platforms. 
{\XRTDAS} is part of the HEAsoft Swift package for the analysis of all
\Swift\ data and 
is distributed by the HEASARC (http://heasarc.gsfc.nasa.gov).
This distribution includes full documentation for {\XRTDAS}.
 
The SDC is also responsible for the standard automated data processing pipeline of the
\Swift\ ground system, which is
routinely run on all \Swift\ data. 
The output of the pipeline is a standard set of data products that are
archived and made publicly available by the on-line services of the HEASARC at NASA/GSFC, the
Italian Swift Archive Center (ISAC, operated jointly by the ASDC and OAB), and the UK
Swift Science Data Centre (UKSSDC) at the University of Leicester. 
{\XRTDAS} can also be used to reprocess low level archival XRT data to
apply  non-standard screening or filtering criteria, or to take into account changes in the calibration files. 

{\XRTDAS} supports the processing of all XRT readout modes. 
The main steps of the XRT standard data processing pipeline are summarized below.
In addition, {\XRTDAS} includes a number of tasks, such as
centroid determination
and exposure map calculation that are not used in the standard
pipeline.  

Steps common to all XRT science readout modes:
\begin{itemize}
\item Correction of the spacecraft attitude file using the information
      obtained with the Telescope Alignment Monitor (TAM)
\item Generation of a filter file containing instrument housekeeping,
      attitude and orbit-related parameters values, to be used for the
      calculation of the time intervals where the events are
      considered acceptable for science data analysis (Good Time
      Intervals, or GTI). 
\end{itemize}

Steps specific to Image Mode processing: 
\begin{itemize}
\item Bias subtraction, removal of calibration sources and of bad pixels. 
\item Calculation of an image in sky coordinates using spacecraft attitude and telescope alignment.
\end{itemize}

Steps specific to Photon Counting mode:
\begin{itemize}
\item Assignment of XRT Grade for the 3x3 charge distribution matrix (see Figure~\ref{fig:PC_grades}) and
       calculation of PHA values.  (Bias subtraction in PC mode is
       performed on-board.)
\item Gain correction for Charge Transfer
      Inefficiency (CTI) to convert X-ray PHA values into Pulse Invariant (PI)
      energies.
\item Transformation from raw to sky coordinates using spacecraft attitude and the telescope alignment.
\item Flagging and screening of events associated with the calibration
      sources, and with bad, hot or flickering pixels.
\item Selection of events based on grade and good time intervals.  
\item Creation of high-level science products (e.g. spectra, lightcurves and images).
\end{itemize}

Steps specific to timing modes (Windowed Timing and Photodiode modes): 
\begin{itemize}
\item Assignment of the photon arrival time based on CCD photon position. For the case of Photodiode modes all photons are assumed to come from the target of the observation. 
\item Flagging of partially exposed and piled-up frames and bias subtraction when not done on board (Photodiode modes only).
\item Event recognition using a 7x1 pixel array, grade assignment
      (see Figure~\ref{fig:timing_grades}), and PHA calculation.
\item Gain correction for Charge Transfer Inefficiency (CTI). 
\item For Windowed Timing mode, which has one-dimension imaging
      capability, transformation from raw detector to sky coordinates
      using spacecraft attitude and the telescope alignment. For
      Photodiode mode, which has no imaging capability, only detector coordinates are calculated.
\item Flagging and screening for events associated with bad columns (Windowed Timing mode only).
\item Selection of events based on grade and Good Time Intervals.  
\item Creation of high-level science products (e.g. spectra,
      lightcurves and images) from the screened event file. Since no
      position information is available in Photodiode mode, no spatial filtering is applied in this case.
\end{itemize}

Images and spectra obtained through TDRSS messages can also be
processed with \XRTDAS.  In this case images can be displayed in sky
coordinates, source fluxes can be estimated and data files can be
converted from their telemetry format into FITS files usable with
standard imaging and spectral analysis packages.

\section{XRT OPERATIONS}
With the exception of occasional manual calibration observations, the
operation of the XRT is completely autonomous and is driven by the
response of the observatory to its on-board observation timeline and
to newly discovered GRBs.  The \Swift\ Mission Operations Center (MOC)
will load pre-planned target (PPT) timelines onto the observatory on a
daily basis or as needed.  New GRBs trigger Automated Targets (ATs),
which are generated on-board by the BAT instrument.  A software
process called the Figure of Merit process controls the observing
timeline and arbitrates between PPTs and ATs according to their
assigned observing priorities.

\begin{figure}
\centering
\includegraphics[bb=30 225 750 750,clip=true,width=\maxfloatwidth,angle=0]{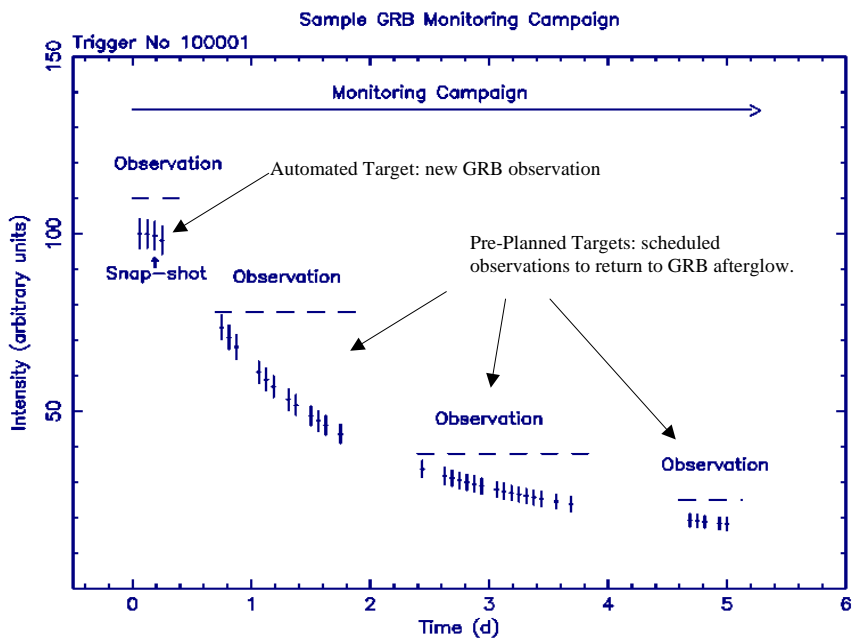}
\caption{Typical \Swift\ observing program for a new GRB.  The observing
  program is broken into several ``observation segments'' (labelled
  ``observations'' on the figure), each lasting 1-2 days, which are 
in turn broken into ``snapshots'' of typical duration 20-30 minutes.  
The first observation segment is scheduled automatically on-board and 
lasts 60,000 s (on-target).  Subsequent observation segments are 
pre-planned on the ground as updates to the mission timeline.}
\label{fig:observations}
\end{figure}

Because the \Swift\ observatory is in low Earth orbit, and because the
two Narrow-Field Instruments (XRT and UVOT) cannot be pointed closer
than 30 degrees to the Earth's limb, there is no ``Continuous Viewing
Zone" for \Swift, and all observations will be broken up by observing
constraint violations that require the spacecraft to slew to a new
target.  This results in observations of any given target that are
broken into segments of (typically) 20-30 minutes duration, which we
refer to as ``snapshots'', with other objects intervening.  This is
illustrated graphically in Figure~\ref{fig:observations}.  Snapshots are grouped together
into ``observation segments'' that cover several days.
A single roll angle will be used for each observation segment. 
In ground processing, each observation segment is processed as a
single data set.

\section{ALIGNMENT}
\subsection{Ground Alignment}
The telescope optical pointing direction or boresight was
established by using a micro-alignment telescope (MAT) mounted on a
special alignment jig to accurately measure the line from the center
of the CCD through the center of the mirror module.  The measured
boresight was transferred to an external alignment reference
cube, using optical transfer flats and an additional autocollimator.
Following alignment, this master alignment cube defines the XRT
instrument axes, with the X-axis parallel to the XRT boresight (+X
towards the target), the Y-axis parallel to the CCD serial register,
and +Z in the CCD parallel readout direction.

\subsection{On-Orbit Alignment}
On-orbit alignment was accomplished through a two-step process.
In the first step, the instrument was pointed at several bright stars.
The positions of the stars were
measured in both the XRT (by optical light leak through the blocking
filter) and the star tracker images, and the star
tracker software parameters were updated to zero out the boresight alignment
so that the star tracker reports its orientation with respect to the
XRT boresight.  Secondary alignment measurements were made by the
XRT, and XRT instrument software parameters were adjusted to zero
out residual offsets to the Star Tracker coordinate system. We
expect on-orbit alignment to be accurate to within 2 arcseconds.

\section{CALIBRATION}
The XRT flight detector and optical blocking filter were calibrated at the
University of Leicester.  The mirrors were calibrated at the Panter
X-ray Calibration Facility operated by MPE in Neuried, Germany.  
End-to-end testing and calibration of the fully assembled
instrument was also performed at the Panter facility in September
2002.  
This end-to-end calibration verified
the instrument point spread function, effective area, and focus.
Results of these calibrations have been reported elsewhere 
(\opencite{9}; \opencite{13}; \opencite{14}) and are incorporated in
the XRT response matrices, which are available from the Swift Science
Center at NASA/Goddard Space Flight Center ({\it http://swiftsc.gsfc.nasa.gov/docs/swift/swiftsc.html}).

\subsection{CCD Selection and Calibration} Four \Swift/XRT CCDs were tested,
characterized, and calibrated at the
University of Leicester. The UL X-ray calibration facility uses a Kevex source and
fluorescent targets, plus an electron-bombardment source with coated
anodes and a monochromator, to generate X-ray emission lines from Boron K$\alpha$
(183 eV) to Arsenic K$\alpha$ (10532 eV). At energies below
1500 eV, a crystal monochromator is used to isolate the desired
emission line. The CCD being tested is mounted on a liquid nitrogen cold finger,
and its temperature may be controlled in the range -40 to -140~C. 
The X-ray beam flux is monitored with a calibrated lithium drifted
silicon detector to measure quantum efficiency.

Selection of potential flight devices was based upon cosmetic defects
(bright and dark pixels and columns), spectral resolution across the
energy band, quantum efficiency and spatial uniformity of the low
energy response. One flight detector and three spare devices were 
calibrated over a range of energies, temperatures and flux rates, with
nearly a million photons  collected on the flight CCD \cite{Beardmore05}.  The
calibration data were used to tune our Monte Carlo CCD model, and
response matrices have been generated for all XRT readout modes
(\opencite{14}; \opencite{Mukerjee05}).

\subsection{Mirrors}  The flight mirrors were calibrated at the
Panter X-ray Calibration
Facility operated by MPE in Neuried, Germany as part of the JET-X calibration in 1996 \cite{Citterio96} and were
stored at the Osservatorio Astronomico di Brera in a dry, inert
atmosphere in a hermetically sealed shipping container for the next
four years.  They were recalibrated in July 2000 at the Panter
facility to check for any changes in performance since the original
calibration.  Analysis of these data confirmed the previous results.

\subsection{End-to-end Calibration}  
\begin{figure}
\centerline{%
\begin{tabular}{cc}
\includegraphics[bb=154 170 650 719,clip=true,width=2.05in,angle=0]{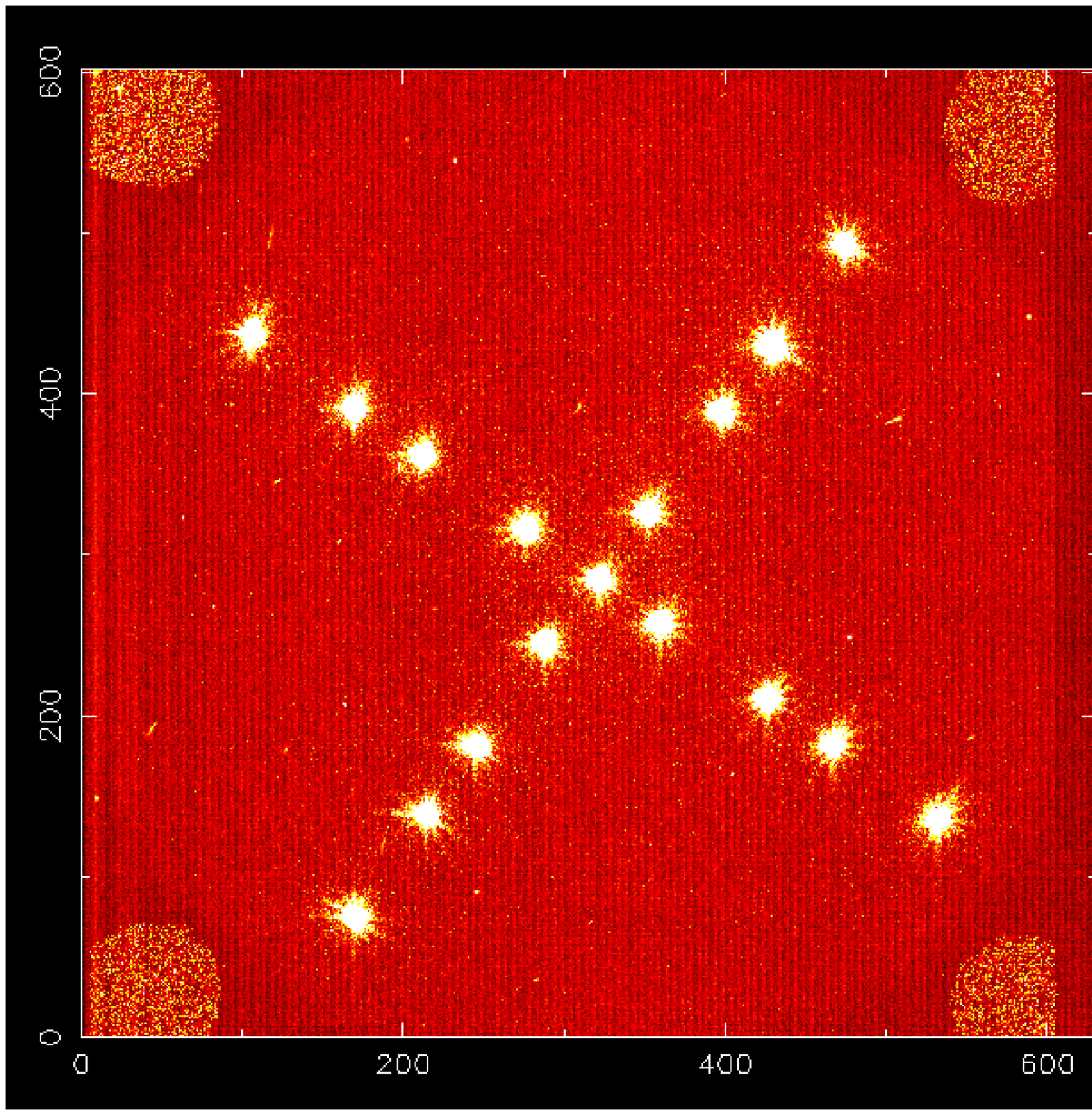}
&
\includegraphics[bb=27 125 570 720,clip=true,width=2.3in,angle=90]{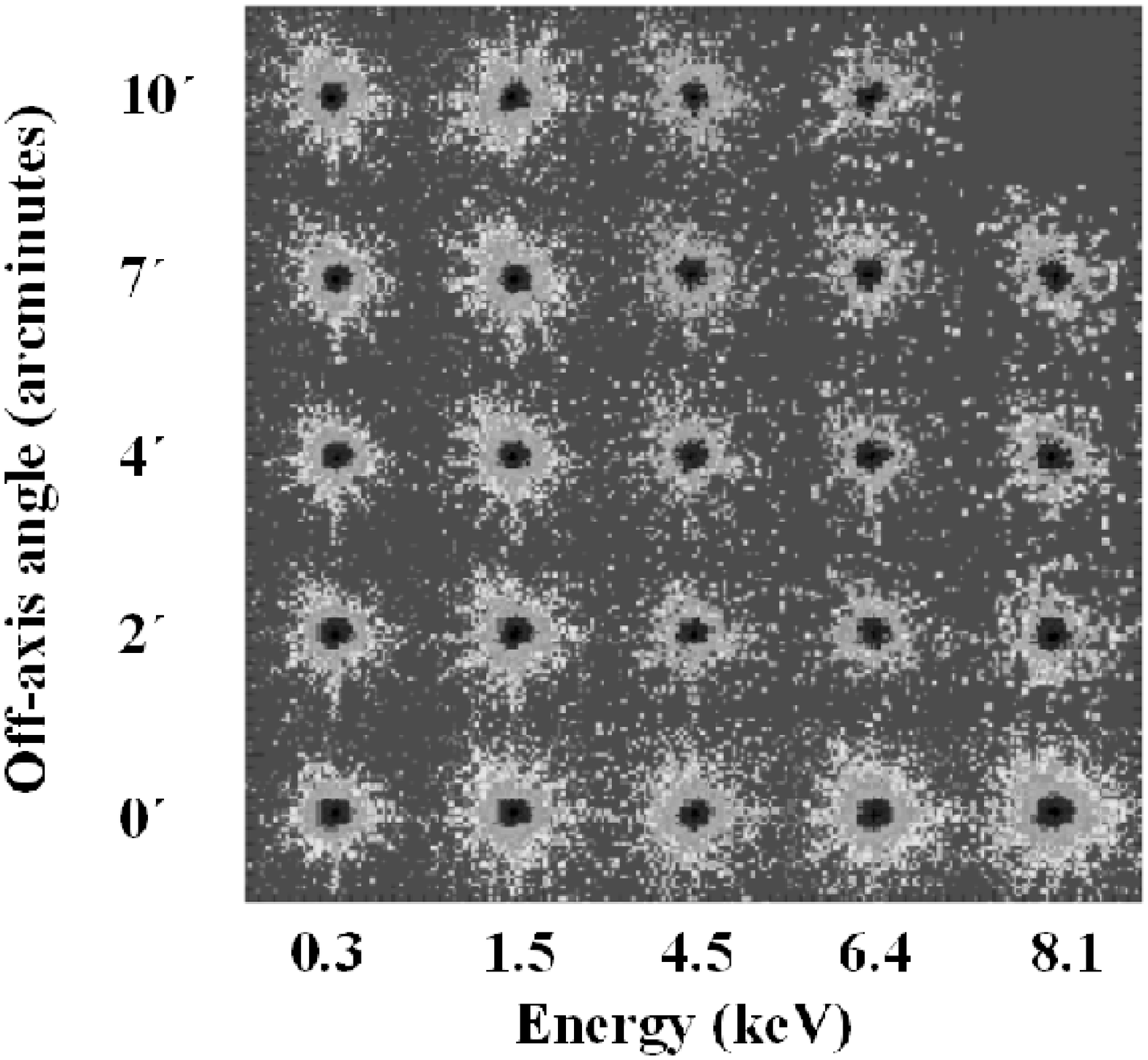} \\
\end{tabular}}
\caption{{\bf Left:} Composite of 1.49 keV calibration images taken at Panter at a variety of
  off-axis angles spanning the XRT field of view.  The image quality
  is quite uniform over the entire field of view.  The large, diffuse
  circles in the corners of the detector are the on-board calibration
  sources.  The dark band on the right-hand side of the image is the
  over-clocked pixels, which are used to measure the baseline level.
{\bf Right:} composite images made at five different energies and five
off-axis angles.  The PSF degrades significantly at high energies due
to increased scattering from the mirror surfaces, but still retains a
sharp core.}
\label{fig:psf2}
\end{figure}

The end-to-end calibration concentrated on four
primary measurements:
\paragraph*{Verification of the focus of the telescope:} The XRT has a
fixed focus.  An important part of the end-to-end calibration was
verification that this focus is accurate within the 1 mm depth of
field of the optics.  A spacer was installed between the aft telescope
tube and the FPCA in order to compensate for the finite source
distance.  The calibration data verified that the instrument is
properly focused (see Figure~\ref{fig:psf2}).

\paragraph*{Point-Spread Function:} Measurement of the on-axis and off-axis PSF of the telescope at
      several energies verified that the mirror mounting has not
      introduced any distortion to the mirrors.  Composite
      images made from these data are shown in Figure~\ref{fig:psf2}.
      These data were fitted to an analytical model that provides an
      excellent approximation of the PSF of the telescope as a
      function of energy and off-axis angle.  For details, we refer
      the reader to \inlinecite{9}.

\paragraph*{Effective Area:} We measured the end-to-end effective area
of the XRT at several energies and off-axis angles.  These data were
used to derive on-axis effective area curves and vignetting functions 
for all of the XRT readout modes \cite{13}.  The results have
been incorporated into an ancillary response file generator program, {\it mkarf}, that creates an
appropriate effective area curve (ARF file) for each observation.

\subsection{In-Flight Calibration} In-flight calibrations have been
performed  during the Performance Verification phase following
instrument turn-on, and will continue as an ongoing activity.  The PV phase
calibration includes the following:
\begin{itemize}
\item Calibration of the XRT PSF as a function of off-axis angle, using
      bright X-ray point sources with hard and soft spectra (PKS
      0312-770, RX J0720.4-3125, GX1+4, RXS J1708-4009).
\item Calibration of the end-to-end instrument effective area, using
      supernova remnants with well-known and
      constant spectra (Crab, Cas A, 2E0102-7217).
\item Calibration of the energy response using bright emission line
      sources (2E0102-7217, Cas A, AB Dor, HD 35850, Capella).
\item Calibration of the timing modes using X-ray pulsars
      (RXS J1708-4009, Crab), cross-calibrated to RXTE.
\item Special calibrations include determination of off-axis stray
      light sensitivity, optical blocking filter transmission,
      boresight alignment, and centroiding accuracy.
\end{itemize}
These calibration observations will be repeated on a periodic basis (roughly once every six months) to monitor any instrument changes with time.

In addition to calibration with astrophysical sources, the XRT will be
calibrated on an ongoing basis using four small on-board $^{55}$Fe sources
that constantly illuminate the corners of the CCD at a low rate.
These events (which are outside the instrument field of view) are tracked as part of
our trend analysis to monitor energy resolution, gain, and charge
transfer efficiency throughout the mission.

Finally, full CCD images are sent to the ground on a regular basis
to monitor the detector for any changes in hot pixels or other
artifacts that could affect the on-board event processing.  On-board
lists of bad rows, columns, or pixels will be updated based on
analysis of these images by the instrument team.

\subsection{Pile-up}
\label{sec:pileup}
The automated operating modes of the XRT are designed to minimize
pileup for most observations.  Photon pileup should be minimal
for source fluxes below 1 Crab.  However, the
earliest observations of the brightest GRB afterglows may occur while the source
flux is more than 10 times brighter than this, and these early data may be
significantly piled-up.  Two strategies have been developed to cope
with pileup in these cases:

\paragraph*{Photodiode Mode:} Pileup in PD mode data will occur for
source fluxes in excess of $\sim 6 \times 10^{-8}$ \flux.  Because pileup changes the
electron distribution in the detector, it is accompanied by a loss of
information and cannot be directly corrected. We are modelling the
effects of pileup for PD mode in order to produce algorithms designed
to allow the user to obtain approximate spectral parameters, even in
the event of severe pileup, following the technique described for
piled-up {\it Chandra} observations of the Crab nebula by \inlinecite{Mori04}.

\paragraph*{Windowed Timing Mode:} In WT mode the CCD is clocked
continuously in the parallel shift direction, so that an image made
from these data consists of a vertical streak across the CCD.
Pileup in WT mode results in a piled-up central streak, surrounded on
both sides by lower count rates resulting from the wings of the PSF.
Very bright sources with severe pile-up can be analyzed in WT mode by
utilizing only the wings of the PSF.  The XRT {\it mkarf} program can
produce the special ARF files required for this type of analysis.

\begin{acknowledgements}
This work is supported at Penn State by NASA contract NAS5-00136; at
the University of Leicester by the Particle Physics and Astronomy
Research Council on grant number PPA/G/S/00524; and at INAF-OAB by funding
from ASI on grant number I/R/309/02.  We gratefully acknowledge the
contributions of dozens of members of the XRT team at PSU, UL, OAB,
GSFC, and our subcontractors, who helped make this instrument
possible.
\end{acknowledgements}


\end{article}

\end{document}